\documentclass[12pt]{article}

\usepackage{amssymb}
\usepackage{a4wide}

\begin{document}
\sloppy

\newcommand{\club}{$\clubsuit \ $}
\newcommand{\mclub}{{\;  \clubsuit \;}}
\newcommand{\mmclub}{$ \clubsuit \clubsuit \ $}
\newcommand{\bul}{\P \ }
\newcommand{\mbul}{\mbox{\P}\ }
\newcommand{\mmbul}{\P\P\ }

\newcommand{\LA}{\Omega}
\newcommand{\hh}{{\cal H}}

\newcommand{\Q}{ {\mathbb Q}\, }
\newcommand{\D}{ {\mathbb D}\, }
\newcommand{\cq}{ {\bf {\cal Q}}}

\newcommand{\zz}{ {\bf Z} }

\newcommand{\vf}{\varphi}
\newcommand{\pa}{\partial}

\newcommand{\re}{{\rm e}}
\newcommand{\ri}{{\rm i}}
\newcommand{\br}{{\bf R}}

\newcommand{\half}{1\over 2}
\newcommand{\hs}{{\cal H}}

\newcommand{\beq}{\begin{equation}} 
\newcommand{\eeq}{\end{equation}} 
\newcommand{\bea}{\begin{eqnarray}} 
\newcommand{\eea}{\end{eqnarray}}
\newcommand{\pr}{\prime}

\newcommand{\ds}{\displaystyle}
 
\newtheorem{theo}{Theorem}[section]
\newtheorem{lem}{Lemma}[section]
\newtheorem{prop}{Proposition}[section]
\newtheorem{cor}{Corollary}[section]
\newtheorem{defin}{Definition}[section] 
 

\newcommand{\nc}{\newcommand}
\nc{\N}{{\rm I\mkern-4.0mu N}}        
\nc{\R}{{\rm I\mkern-4.0mu R}}
\nc{\C}{{\rm I\mkern-4.0mu C}}
\nc{\Z}{{\sf Z\mkern-6.5mu Z}}
\nc{\Id}{{\mathchoice {\rm 1\mskip-4mu l} {\rm 1\mskip-4mu l}
{\rm 1\mskip-4.5mu l} {\rm 1\mskip-5mu l}}}

\hfill LYCEN  2004-03
\vskip 0.07truecm
\hfill January 2004 
 
\thispagestyle{empty}

 
\bigskip
\begin{center}
{\bf \LARGE{Are $N=1$ and $N=2$}}
\end{center}
\begin{center}
{\bf \LARGE{supersymmetric quantum mechanics}}
\end{center}
\begin{center}
{\bf \LARGE{equivalent?}}
\end{center}

\bigskip

\centerline{{\bf Monique Combescure, 
Fran\c cois Gieres and Maurice Kibler}}

\bigskip
\centerline{{\it Institut de Physique Nucl\'eaire de Lyon}}  
\centerline{{\it IN2P3-CNRS et Universit\'e Claude Bernard}}
\centerline{{\it 4, rue Enrico Fermi}} 
\centerline{{\it F-69622 Villeurbanne Cedex}}

\bigskip
\bigskip
\bigskip

\begin{abstract}
After recalling  
different formulations of the definition of 
supersymmetric quantum mechanics given in the literature, 
we discuss 
the relationships between them
in order to 
provide an answer to the question raised 
in the title.
\end{abstract}

\newpage

\thispagestyle{empty}
\tableofcontents

\newpage
\setcounter{page}{1}

\section{Introduction}


Supersymmetric quantum mechanics (SUSYQM) was introduced 
more than two decades ago by 
Nicolai~\cite{nic} and 
Witten~\cite{w1}. 
In its simplest form, it consists of the study of 
quantum mechanical systems 
which are described by a Hamiltonian operator of the form 
$H=\Q ^2$ acting on a Hilbert space $\hs$ which admits a 
$\zz_2$-grading, i.e.,  $\hs$ has the form of a 
direct sum: $\hs = \hs_b \oplus
\hs_f$. 

 The aim of the present letter is to elucidate 
the precise relationship 
 between different formulations 
 of the definition for SUSYQM
  which have been considered in the 
literature~\cite{nic}-\cite{bs}.
   To start with, we will list these different defining 
 relations as well as the results to be established in
the present work.

 \section{Definitions and summary of results} 
 
 The common starting point is a {\em quantum mechanical system} 
 $( \hs , H)$ 
 characterized by a self-adjoint operator 
$H \neq 0$ 
 (the {\em Hamiltonian} or 
 {\em energy}
 of the system) acting on a
 complex separable Hilbert space $\hs$ (the {\em state space}). 
As usual, the  commutator and anticommutator of two
operators $A$ and
$B$ are denoted by $[A,B]$ and 
$\{ A,B \}$,
respectively. 
All operators to be considered are linear and the
adjoint of an operator $A$ is denoted by $A^{\dagger}$. 
Our concern will primarily be of algebraic nature
and we leave it to the mathematically minded reader 
to supplement the relevant analytical details like 
domains of definition for operators, 
proper characterization of the anti-commutativity for 
unbounded self-adjoint operators, etc.~\cite{vasil}.

\subsection{Definitions}\label{def}

 \begin{enumerate}\item[]
{\bf Definition 1 : } 
    {\it 
    The quantum mechanical system  $( \hs , H)$ is 
    called {\em supersymmetric}
    if there exists a finite number of self-adjoint operators
    $Q_1,\dots ,Q_N$  on $\hs$  
   such that 
    \beq 
    \label{qi}
    \{ Q_i , Q_j \} = 2 \delta_{ij} H 
  \qquad \quad {\rm for} \ \; i,j \in \{1,\dots ,N \}    
 \ .
 \eeq
 The operators $Q_1,\dots ,Q_N$ are called {\em supercharges}
(or {\em supersymmetry generators}).
} 
 \end{enumerate}
 From relations (\ref{qi}), it follows that the supercharges are
 conserved, i.e., that they commute 
    with the Hamiltonian:
${[ H, Q_i ]} = 0$ for $i \in \{1,\dots ,N \}$.
Since the latter relation 
also means that the Hamiltonian is 
invariant under the transformations generated by the $Q_i$'s, 
the operator $H$ is called a {\em supersymmetric Hamiltonian}
or {\em super-Hamiltonian}.
With supersymmetric field theories 
 \cite{wess} in mind, the algebra (\ref{qi}) with $N$ supercharges
 is usually qualified 
 as {\em $N$-extended supersymmetry algebra}.

 \begin{enumerate}\item[]
{\bf Definition 2 : } 
    {\it 
    The quantum mechanical system $( \hs , H)$ is 
    called {\em supersymmetric}
   if there exists a finite number of non self-adjoint operators
    $q_1,\dots ,q_M$  on $\hs$  
   such that 
       \beq 
       \label{qqd}
    \{ q_i , q_j ^{\dagger} \} = 2   \delta_{ij} H 
  \ , \qquad 
     \{ q_i,q_j \} = 0 
   \qquad \quad {\rm for} \ \; i,j \in \{1,\dots ,M \}       \ .
\eeq
 The operators $q_1, \dots , q_M$ are called {\em 
 complex supercharges}. 
 }
 \end{enumerate}
  It follows that 
  $\{ q_i^{\dagger} , q_j^{\dagger} \} =0$ and that 
 $[H,q_i] =0= [H, q_i^{\dagger}]$  
 for $ i,j \in \{1,\dots ,M \} $.
Note that the operators    $q_1,\dots ,q_M$ satisfying (\ref{qqd})
cannot be self-adjoint, otherwise $H=0$ (contrary to our 
assumption $H \neq 0$).

 \begin{enumerate}\item[]
{\bf Definition 3 : }
    {\it 
    The quantum mechanical system $( \hs , H)$ is said to be 
    {\em supersymmetric} if there exists a finite number of self-adjoint
    operators ${\bf \Q} _1,\dots , {\bf \Q} _n$  
(called {\em supercharges}) 
    as well as a bounded self-adjoint operator $K$ 
    (called {\em involution}),
all of which operators act on 
    $\hs$  and  satisfy 
    \bea 
    K^2 \!\!\! &=& \!\!\!  \Id 
  \nonumber   \\
    { \{K,\Q_i \} } \!\!\!  &=& \!\!\!  0
  \qquad \, {\rm for} \ \; i \in \{1,\dots ,n \}   
 \label{kq}
 \eea
and 
    \beq
 \label{qqh}
 \{ \Q _i , \Q _j \} = 2 \delta_{ij} H 
   \qquad \quad {\rm for} \ \; i,j \in \{1,\dots ,n \}       
   \ .   
\eeq
}
 \end{enumerate}
The operator $K$ is also referred 
    to as {\em Klein operator}, 
    {\em chirality operator}, 
    {\em fermion number operator}, 
    {\em Witten parity operator} or 
{\em  $\zz_2$-grading operator}.
 Since $K$ is a bounded operator, it can be (and it will be) assumed to be
 defined  on the entire space $\hs$. 
Note that 
the choice 
$K= \pm \Id $ implies
 the trivial solution 
$\Q_1 = \dots = \Q_n =H =0$ 
that we excluded. 
Definition 3 also implies $[H,\Q_i]=0$ for all values of $i$. 

 \begin{enumerate}\item[]
{\bf Definition 4 : }
{\it 
The present definition (involving 
$m$ complex supercharges ${\bf q}_1, \dots ,{\bf q}_m$)
is the same as Definition~2  supplemented with an involution $K$.
}
\end{enumerate}

  Obviously, 
Definitions 3 and 4 are nothing but Definitions 1 and 2, respectively,
supplemented with 
the operator $K$. The crucial question is whether the existence
  of this operator already follows from the existence 
  and properties of the supercharges, i.e., whether 
$K$ necessarily
  represents a function of the supercharges or whether it is
an extra independent input. 
  
   In our study of the relationship between
  the given definitions we will concentrate on
  the most popular special cases which we now summarize
  for later reference.

\subsection{The most important special cases}\label{special}   
   The simplest and most studied supersymmetric systems are, 
   respectively, given by 
      the following cases: \\
 $\ast$ $N=2$ {\em in Definition 1,} i.e., 
 \beq 
    \label{qi2}
Q_1^2 = Q_2^2 =H 
 \, , \qquad 
 \{ Q_1 , Q_2 \} = 0
\, .
\eeq 
 $\ast$ $M=1$ {\em in Definition 2}, i.e.,  with the notation 
$q \equiv q_1$:
 \beq 
   \label{qi3}
   \{ q , q ^{\dagger} \} = 2  H 
  \, , \qquad 
     q^2 = 0 \, .
\eeq
$\ast$ $n=1$ {\em in Definition 3}, i.e., 
with the notation $\Q \equiv \Q_1$:
\beq   
  \label{qi4}
  K^2 = \Id \, , \quad  
 { \{K, \Q \} } = 0 
\, , \quad  \Q ^2 =H \, .
  \eeq 
$\ast$ $n=2$ {\em in Definition 3}, i.e., 
\beq   
  \label{qi7}
  K^2 = \Id \, , \quad  
 { \{K, \Q_1 \} } = { \{K, \Q_2 \} } =0
\, , \quad  \Q_1 ^2 =  \Q_2 ^2 = H
 \, , \quad  
 \{ \Q_1 , \Q_2 \} = 0 \, .
\eeq 
$\ast$ $m=1$ {\em in Definition 4}, 
i.e., one has a non self-adjoint operator ${\bf q}$ satisfying 
\beq   
  \label{qi5}
K^2 = \Id \ , \quad   
 { \{K, {\bf q} \} } = 0 \ , \quad 
 \{ {\bf q}, {\bf q} ^{\dagger} \} = 2  H 
\ , \quad  {\bf q} ^2 =0 \, .
  \eeq 

\subsection{Summary of results}\label{summ} 
   
For the sake of clarity,  
we summarize the relationships
between the special cases (\ref{qi2})-(\ref{qi5})
which are going to be established in the sequel.

\bigskip

\noindent 
\underline{{\bf A}}.
By writing 
$q = {1 \over \sqrt 2} \, (Q_1 + \ri Q_2)$
and ${\bf q} = {1 \over \sqrt 2} \, (\Q_1 + \ri \Q_2)$, one 
checks the equivalence of (\ref{qi2}) and (\ref{qi3})
and the equivalence of (\ref{qi7}) 
and (\ref{qi5}).

\medskip

\noindent 
\underline{{\bf B}}.
Relations (\ref{qi4}) imply that the operator 
$\Q ^{\prime} 
= \pm \ri K\Q$ 
represents a second 
supercharge, i.e., remarkably enough, 
 $n=1$ implies $n=2$.

\medskip

\noindent 
\underline{{\bf C}}.
The converse of (B) 
also holds, i.e., the two supercharges 
defining a $n=2$ supersymmetric system are related 
by $\Q_2 = \pm \ri K \Q_1$.

\medskip

\noindent 
\underline{{\bf D}}.
From relations (\ref{qi2}), one can deduce the
existence of an involution operator $K$ such that relations
(\ref{qi7}), or equivalently (\ref{qi4}) 
or (\ref{qi5}),
hold.

\medskip

\noindent 
\underline{{\bf E}}. 
From relations (\ref{qi5}), one concludes that 
$K, {\bf q}$ and $H$ have the general form 
\beq
\label{witf}
K =  \left[
\begin{array}{cc}
    \Id &0 \\
   0& -\Id
   \end{array}
\right] 
\; , \qquad 
{\bf q} = \sqrt{2} \, \left[
\begin{array}{cc}
   0 & A ^{\dagger} \\
   0&0
   \end{array}
\right]
\; , \qquad 
H = \left[
\begin{array}{cc}
A ^{\dagger} A & 0  \\
   0&A A ^{\dagger}
   \end{array}
\right]
\; ,
\eeq
where $A$ is a linear operator. 
In the literature, these expressions for 
the complex supercharge and for the associated Hamiltonian 
(eventually with 
a specific choice of $A$ in terms of the operators of 
position and momentum)
are referred to as {\em Witten's model} of SUSYQM. 

\bigskip 

By combining the previous results, 
we conclude that the sets of  relations
(\ref{qi2})-(\ref{qi5}) are equivalent and that 
every supersymmetric Hamiltonian 
satisfying any one of these sets of  relations
can be cast into  the form (\ref{witf}). 
In other words, the latter expressions do not simply 
describe a  specific model of
SUSYQM (with one complex or two real supercharges),
but they represent its most general 
form\footnote{To be more precise, (\ref{witf}) 
is the most general form up to 
redefinitions of the supercharges leaving the Hamiltonian 
invariant - see Subsection~\ref{unique} below. In particular,  application of unitary transformations may lead to
more complicated expressions for the operators.}.

\bigskip

The proof of statements 
C and E will be provided  
in Subsection~\ref{supinv} and the one of D
in Section~\ref{relations}.
The result B readily follows from the properties  
of the involution $K$~\cite{bs}.
The proof of A is as follows.
Note that 
$q = {1 \over \sqrt 2} \, (Q_1 + \ri Q_2)$
implies 
$ q^{\dagger} = {1 \over \sqrt{2} } 
  \left(Q_{1} - \ri Q_{2} \right)$ and that 
these expressions are equivalent to 
$Q_{1} = {1 \over \sqrt{2} } ( q +  q^{\dagger}) , \, 
Q_{2}  = {-\ri \over \sqrt{2} } (q -  q^{\dagger})$.
From 
\begin{eqnarray*}
   0 \!\!\!&=& \!\!\! q^2 = {1 \over 2} \, 
   \left( Q_1^2 - Q_2^2 + \ri \{ Q_1, Q_2 \} \right)
   \quad \Longleftrightarrow \quad 
 Q_1^2 = Q_2^2  \ \; {\rm and}  \ \;
 \{ Q_1, Q_2 \} =0
 \\
2H \!\!\!&=& \!\!\! \{ q , q^{\dagger} \} 
= Q_1^2 + Q_2^2
\; , 
   \end{eqnarray*}
one concludes that 
$H = Q_1^2 = Q_2^2$ with $ \{ Q_1, Q_2 \} =0$
and vice versa.

   \subsection{Non-uniqueness of supercharges
      and extra symmetries}\label{unique}

Suppose $(\hs , H)$ represents 
a supersymmetric system in the sense
of equation (\ref{qi2}) with supercharges $Q_1$ and $Q_2$.
Then, $H$ can be expressed in a completely symmetric way 
in terms of the supercharges: 
$H = {1 \over 2} (Q_1^2 + Q_2^2)$. 
From this expression, it is clear that 
the charges $Q_1$ and $Q_2$ are not unique: 
the reparametrization
$Q^{\pr} _i = \sum_{j=1}^2 a_{ij} Q_j$, 
where the matrix $A \equiv ( a_{ij})$ 
describes a real 
orthogonal transformation (i.e., $A \in {\cal O}(2)$),  
leaves the defining relations of the supersymmetric system and, in particular,
the Hamiltonian invariant. 
Thus, $Q^{\pr}_1$ and
$Q^{\pr}_2$ represent an equivalent collection of
supercharges for the given supersymmetric system and
a supersymmetric Hamiltonian 
admits a larger invariance than supersymmetry 
since it is automatically 
{\em invariant
under a rotation in the $(Q_1, Q_2)$-space}. 

Similarly, in equation (\ref{qi7}), the 
supercharges 
$\Q_1$ and $\Q_2$
can be transformed by a matrix of $A \in {\cal O}(2)$
and, in (\ref{qi3}) or (\ref{qi5}), 
the complex supercharge 
can be changed by a phase factor 
$\lambda \in {\cal U}(1)$.

\section{Consequences of the definitions}\label{elementary}

In this section, 
we show that Definition 3 with $n=1$,
i.e., equations (\ref{qi4}), 
imply the characteristic features that are generally associated 
with SUSYQM. 
Thereafter, we discuss the general form of 
such supersymmetric systems
and we illustrate the results by 
two simple examples.

\subsection{Characteristic features of supersymmetric systems}\label{charac}
  
We outline the consequences of Definition 3 
for a single supercharge $\Q_1 \equiv \Q$
by expanding on the brief
discussion presented in Reference~\cite{bs}. 
The inner product on the Hilbert space 
$\hs$ will be denoted by  $\langle \cdot , \cdot \rangle$ 
and the induced norm 
by $\| \cdot \|$.
As usual the restriction of an operator $A$ on $\hs$ to a 
subspace ${\cal D} \subset \hs$ is written as 
$A \!  \upharpoonright \! {\cal D}$. 

Since $\Q$ is self-adjoint and $H =\Q^2$, we have 
$H \geq 0$ by virtue of 
 \[
 \langle \varphi , H\varphi \rangle =
\langle \Q \varphi , \Q \varphi \rangle 
= \| \Q \varphi \|^2 \geq 0
\quad 
\mbox{for any} \ \, \varphi \in \hs
\, .
\]
Thus, a supersymmetric Hamiltonian necessarily has 
a {\em nonnegative spectrum}.
As mentioned already,  $H =\Q^2$ implies $[ H , \Q ] = 0$.

From $K^2 = \Id$, it follows that the involution $K$
only admits $\pm 1$ as eigenvalues. Henceforth, $K$ induces a 
direct sum decomposition of the Hilbert space 
$\hs$ : if $\varphi \in \hs$, then 
\begin{eqnarray}
\label{dsd}
\varphi \!\!\! & = & \!\!\!  {1\over 2} \, ( \varphi +K\varphi ) +
{1\over 2} \, ( \varphi -K\varphi )
\\
\!\!\! & \equiv & \!\!\! \varphi_b + \varphi_f
\, .
\nonumber 
\end{eqnarray}
In other words,
\begin{equation}
\label{hsd}
\hs = \hs_b \oplus  \hs_f
\qquad 
{\rm with} \ \
\left\{ 
\begin{array}{l}
 \hs_b  =  \{ \varphi \in \hs \, \mid \, K\varphi =+ \varphi \}
\\
\hs_f  =  \{ \varphi \in \hs \, \mid \, K\varphi =- \varphi \}
\, .  
\end{array} 
\right. 
\end{equation}
Since $K \neq \pm \Id$, the subspaces $\hs_b$ and $\hs_f$ are 
non-trivial, i.e., different from $\hs$ and 
$\{ 0 \}$.
Motivated by the r\^ole played by the operators $\Q$ in particle 
physics, the vectors belonging to  $\hs_b$ and $\hs_f$
are called, respectively, {\em bosonic} (or {\em even})
and {\em fermionic} (or {\em odd}) 
vectors\footnote{One also says that $\hs$ 
is a {\em $\zz_2$-graded}  Hilbert space
with a {\em fixed  parity}~\cite{berezin}.}.
In the present context, this terminology only expresses
the dichotomy introduced into the theory by the 
involution $K$: the precise physical interpretation 
depends on the example under consideration.

It is convenient to introduce a matrix notation 
for the vectors belonging to the direct sum (\ref{hsd}):
rather than writing 
$
\varphi = (\varphi_b, 0) + (0,  \varphi_f ) \, , 
$
we will use the matrix notation
\[
\varphi =
\left[
\begin{array}{c}
\varphi_b \\
0
\end{array}
\right]
+ \left[
\begin{array}{c}
0 \\
\varphi_f
\end{array}
\right]
= 
\left[
\begin{array}{c}
\varphi_b \\
\varphi_f
\end{array}
\right]
\, .
\]
With this notation for the vectors, 
the operator $K$ reads as 
\beq
\label{fnr} 
K= \left[
\begin{array}{cc}
\Id _b & 0 \\
0 & - \Id_f
\end{array}
\right]
\ ,
\eeq
where $\Id _b$
denotes the restriction of the 
identity operator to the subspace ${\hs_b}$ of $\hs$, and 
analogously for $\Id_f$.

The involution $K$ 
not only induces a decomposition of the state space 
$\hs$, but also 
of the algebra of operators acting on $\hs$. In fact, let 
\[
 M = \left[
\begin{array}{cc}
A & B \\
C & D
\end{array}
\right]
\]
denote a generic operator acting on $\hs =\hs_b \oplus \hs_f$. 
Then
\begin{equation}
\label{pa}
[K,M] =0 \qquad \Longleftrightarrow \qquad
 M= \left[
\begin{array}{cc}
A & 0 \\
0 & D
\end{array}
\right]
\end{equation}
and 
\begin{equation}
\label{odd}
\ \ \{ K,M \} =0 \qquad \Longleftrightarrow \qquad
 M= \left[
\begin{array}{cc}
0 & B \\
C & 0
\end{array}
\right]
\ .
\end{equation}
In analogy with the terminology introduced for the state vectors,
the operators commuting with the involution $K$ are called  
{\em bosonic} or {\em even operators} while those anticommuting 
with $K$ are referred to as {\em fermionic} 
or {\em odd operators}.

 Since $\Q$ is self-adjoint 
 and anticommutes with $K$, the result (\ref{odd}) 
 applied to $M= \Q$ implies that 
 \begin{equation}
\label{su}
\Q = \left[
\begin{array}{cc}
0 & A^{\dagger} \\
A & 0
\end{array}
\right]
\ ,
\end{equation}
 where $A$ is a linear operator. Let us now apply $\Q$
 to a vector $\varphi \in \hs$ :
 \[
\Q \varphi
 = \left[
\begin{array}{cc}
0 & A ^{\dagger} \\
A  & 0
\end{array}
\right]
\left[
\begin{array}{c}
\varphi_b \\ \varphi_f
\end{array}
\right]
=
\left[
\begin{array}{c}
A ^{\dagger} \varphi_f \\ A  \varphi_b
\end{array}
\right]
\ .
\]
Since the resulting vector again belongs 
to the space $\hs_b \oplus \hs_f$,
we have 
\begin{eqnarray}
\label{exc}
\Q & : & \hs_b \ \to \ \hs_f
\\
\Q & : & \hs_f \ \to \ \hs_b
\ ,
\nonumber
\end{eqnarray}
which means that $\Q$ {\em exchanges bosonic and fermionic states.} 
It is precisely this fundamental property of $\Q$
which is at the origin of the terminology `supersymmetry' operator.

By virtue of 
$H = \Q^2 $
and (\ref{su}), 
the Hamiltonian $H$ has the form
\begin{equation}
\label{ferm}
H  
= \left[
\begin{array}{cc}
A^{\dagger} A & 0 \\
0& A A^{\dagger} \\
\end{array}
\right]
\equiv 
\left[
\begin{array}{cc}
H_+ & 0 \\
0& H_- \\
\end{array}
\right]
\, ,
\end{equation}
with $H_+ : \hs_b \to \hs_b$ and 
$H_- : \hs_f \to \hs_f$.

We remark that expressions (\ref{fnr}),(\ref{su}) and (\ref{ferm})
are known as the {\em standard} or 
{\em fermion number representation}
of SUSYQM. Equivalent though more complicated expressions 
can be obtained by applying a unitary transformation to all
of these operators.

To conclude, we come to the fundamental spectral property of every
supersymmetric system. Suppose\footnote{Here, the real number $E$ 
belongs to the discrete spectrum of $H$ 
if $\varphi$ is an element of the Hilbert space $\hs$
(or, more precisely, if it is an element of the domain ${\cal D} (H) \subset
\hs$ of the operator $H$).  
It belongs to the  continuous spectrum of $H$ 
if  $\varphi$ represents
a weak (distributional)  solution of the eigenvalue equation.} 
\[
H\varphi = E \varphi
\qquad {\rm with} \qquad E>0
\ .
\]
By applying the operator $\Q$ to this relation and using 
$[H,\Q ]=0$, we find 
\[
H(\Q \varphi)  = E (\Q  \varphi )
\ .
\]
Hence, if $\varphi$ is an eigenstate of the Hamiltonian $H$,
then $\Q \varphi$ also represents an eigenstate of $H$
associated to the same eigenvalue $E>0$.
(This argument is not valid 
for $E=0$ : the relation  $H\varphi =0$ infers 
$
0= \langle \varphi, H\varphi \rangle =
\| \Q \varphi \|^2$, 
therefore $\Q \varphi =0$ is  the null vector which 
is not an eigenvector by definition.)

According to (\ref{exc}), $\varphi \in \hs_b$ (resp.
$\hs_f$) implies  
$\Q \varphi \in \hs_f$ (resp. $\hs_b$). 
Thus, we have derived the following fundamental property of 
a quantum mechanical system which is supersymmetric
in the sense of Definition 3. 
\begin{theo}[Degeneracy structure of a supersymmetric system]
For a $n=1$
supersymmetric system, the non vanishing 
eigenvalues of the Hamiltonian admit the same number of 
bosonic and fermionic eigenvectors.
\end{theo}
In other words, 
the partner Hamiltonians 
 $H \!  \upharpoonright \!  \hs_b$ and  
 $H \!  \upharpoonright \!  \hs_f$ are 
{\em isospectral}, 
except possibly 
for the eigenvalue zero.
For later reference, we recall that 
the difference 
between the number of bosonic and fermionic 
states of zero energy is known as 
the {\em Witten} or 
{\em supersymmetric index of} $H$~\cite{w1,bs,tha}: 
\begin{eqnarray}
\label{sind}
{\rm ind_S} \, H 
\!\!\! &=& \!\!\! 
{\rm dim} \, {\rm ker} \, \left[ H \!  \upharpoonright \!  \hs_b \right]  -
{\rm dim} \, {\rm ker} \, \left[ H \!  \upharpoonright \!  \hs_f \right]
\\
\!\!\! &=& \!\!\! 
{\rm dim} \, {\rm ker} \, A - 
{\rm dim} \, {\rm ker} \, A^{\dagger} 
\, . 
\nonumber 
\end{eqnarray}
Here, `${\rm ker}$' denotes the kernel and 
$A$ is the operator defining 
the supercharge $\Q$ according to (\ref{su}). 
We note that expression (\ref{sind})  
is only well defined if $\Q$ has some extra 
properties 
like being of Fredholm type, i.e., if the eigenvalue
$0$ of $\Q$ has finite multiplicity~\cite{tha}. 

To conclude, we note that 
in the physically or mathematically interesting applications,
the Hilbert spaces $\hs, \, \hs_b$ and $\hs_f$
are of infinite dimension so that we have the isomorphism
\beq
\label{idfy}
\hs_b \oplus  \hs_f 
\cong 
{\bf C}^2 \otimes \hs_{bf}
\qquad {\rm with} \ \; 
\hs_{bf} \cong \hs_b  \cong \hs_f
\, .
\eeq
The involution $K$ then takes the form 
\beq
\label{kdef}
K = \sigma_3 \otimes \Id 
\qquad {\rm with} \ \; 
\sigma_3 
= \left[
\begin{array}{cc}
 1& 0 \\
0 & -1
\end{array}
\right]
\, , 
\eeq
which is simply written as $K = \sigma_3$  in most of
the literature.
In this setting, 
the $(2 \times 2)$-matrix format of the supercharge 
(\ref{su}) or of the involution (\ref{kdef}) 
can also be expressed 
in terms of the so-called 
{\em fermionic creation and annihilation operators}
which act on the Hilbert space ${\bf C}^2$ and satisfy
canonical anticommutation relations: we have 
\beq
\label{qfca}
\Q = f^{\dagger} \otimes A + f \otimes A ^{\dagger} 
\, , \qquad {\rm with} 
\ \; 
f = 
\left[
\begin{array}{cc}
   0 & 1 \\
   0 & 0 
\end{array}
\right]
\eeq
as well as 
$\sigma_3 = [ f , f^{\dagger} ]$, i.e., 
\beq
K = [ f , f^{\dagger} ] \otimes \Id 
\, , 
\eeq
where
\beq
\label{fca}
\{ f , f^{\dagger} \} = \Id_2 
\, , \qquad
\{ f , f \} = 0
\, .
\eeq
The fact that $\Q$ is linear in the fermionic operators
$f$ and
$f^{\dagger}$ (acting on ${\bf C}^2$)
reflects the fact that $\Q$ is fermionic 
(odd) with respect to the ${\bf Z}_2$-grading on $\hs$.


\subsection{General form of a $n=1$ (or $n=2$) 
supersymmetric system}\label{gfn1}

\subsubsection{Supercharges and involution}\label{supinv} 


Let us consider a $n=1$ supersymmetric system.
By virtue of (\ref{su}), the supercharge
has the general form 
$\Q \equiv \Q_1  = 
\left[ 
\begin{array}{cc} 
0 & A_1^{\dagger} \\
A_1 & 0
\end{array} \right]$. 
We can decompose the operator $A_1$ 
according to $A_1 = a_1 + \ri a_2$ where the 
self-adjoint operators $ a_1$ and $a_2$  
represent 
the Hermitean and anti-Hermitean 
(real and imaginary) parts of $A_1$.
Thus, we have 
\beq
\label{q1}
\Q_1 = 
\left[ 
\begin{array}{cc} 
0 & a_1 - \ri a_2 \\
 a_1 + \ri a_2 & 0
\end{array}
\right]
\, .
\eeq  

Given the operators $a_1$ and
$a_2$, one can find 
a second supercharge of the same form, 
$\Q_2 \equiv 
\left[ 
\begin{array}{cc} 
0 & b_1 - \ri b_2 \\
 b_1 + \ri b_2 & 0
\end{array}
\right]$,  
which is `normalized' in the sense that $\Q_2^2 = \Q_1^2 $ 
and which is `orthogonal' to  $\Q_1$ 
in the sense that $\{ \Q_1 , \Q_2 \} =0$: 
this supercharge is determined up to a global sign 
and given by 
\beq
\label{q2}
\Q_2 = 
\left[ 
\begin{array}{cc} 
0 & -a_2 - \ri a_1 \\
 -a_2 + \ri a_1 & 0
\end{array}
\right]
\, .
\eeq 
In other words, 
$(b_1, b_2)= \pm (-a_2, a_1)$.

One immediately verifies that 
\beq
\label{remres}
\Q_2 = -\ri \, K \Q_1 \, , 
\qquad \mbox{or equivalently,} \ \ 
\Q_1 = +\ri \, K \Q_2
\, , 
\eeq 
hence we have the following general result:
{\em every $n=2$ supersymmetric system is 
of the form 
(\ref{remres}), 
i.e., $n=1$ is equivalent to $n=2$}.
The associated {\em complex supercharge} takes the
simple and well known form 
which was put forward in 
equation (\ref{witf}):
\beq
\label{wel}
{\bf q} \equiv \ds{1 \over \sqrt{2}} \, (\Q_1 + \ri \Q_2) 
= \sqrt{2} \, 
\left[
\begin{array}{cc}
   0 & A_1^{\dagger} \\
   0 & 0 
\end{array}
\right]
\, .
\eeq

If $\hs, \hs_b$ and $\hs_f$ are of infinite dimension,
the operator $A_1$ acts on  
$\hs_{bf} \cong \hs_b \cong \hs_f$
and we can rewrite expressions (\ref{q1}),(\ref{q2}) as 
\begin{equation}
\label{pauli}
\Q_1 = 
 \sigma_1 \otimes a_1 +  \sigma_2 \otimes a_2
\, , \qquad 
\Q_2  =
\sigma_2 \otimes a_1 - \sigma_1 \otimes a_2
\, .
\nonumber 
\end{equation}
Here, $\sigma_1$ and 
$\sigma_2$ are the Pauli matrices 
which represent a
basis of complex Hermitean $(2 \times 2)$-matrices 
anticommuting with $\sigma_3$. (Note that 
the operators (\ref{pauli}) anticommute with $K = \sigma_3 \otimes \Id$.)
These Hermitean matrices generate the Clifford algebra 
associated to the Euclidean
metric in 
a $2$-dimensional vector space,
\beq
\label{clif}
\{ \sigma_{\alpha} , \sigma_{\beta} \} 
=2 \delta_{\alpha \beta} \Id 
\quad \qquad 
(\, \alpha , \beta = 1, 2 \, ) 
\, .
\eeq  

Following the practice of quantum mechanics, 
we can combine the generators $\sigma_1$ and 
$\sigma_2$
satisfying the relation (\ref{clif}) into a  
fermionic annihilation operator
\beq
f \equiv {1 \over 2} (\sigma_1 + \ri \sigma _2 ) = 
\left[
\begin{array}{cc}
   0 & 1 \\
   0 & 0 
\end{array}
\right]
\, ,
\eeq
the latter acting on the Hilbert space ${\bf C}^2$ and satisfying
the canonical anticommutation relations (\ref{fca}).
The general form of supercharges for 
a $n=2$ supersymmetric system, 
as given by (\ref{pauli}), 
then reads as
\beq
\label{qfop}
\begin{array}{ll}
&
\Q_i = f^{\dagger} \otimes A_i + f \otimes A_i ^{\dagger} 
\\
{\rm with} \ \; & 
A_1 \equiv a_1 + \ri a_2 \, , \ 
A_2 \equiv -a_2 + \ri a_1 \, .
\end{array} 
\eeq
The associated complex supercharge is given by 
${\bf q} = \sqrt{2} \, f \otimes A_1^{\dagger}$ 
and the {\em Hamiltonian} 
$H = \Q_1^2=\Q_2^2$ 
takes the form  
\bea
\nonumber 
H 
\!\!\! &=& \!\!\!
\{ f, f^{\dagger} \} \otimes (a_1^2 + a_2^2) + 
 [f, f^{\dagger}] \otimes 
\ri \, [a_1, a_2 ] \\
\!\!\! &=& \!\!\! \Id_2 \otimes (a_1^2 + a_2^2) + \sigma_3 \otimes 
\ri \, [a_1, a_2 ]
\, . 
\label{typ}
\eea


\subsubsection{Examples} 

Many interesting Hamiltonians are supersymmetric,
e.g. see references~\cite{nic}-\cite{bs}. 
The prototype examples are the spin-${1 \over 2}$ particle
in a one-dimensional superpotential or in a constant 
two-dimensional magnetic field. 
A simple and important example of a more mathematical nature is the 
Laplace-Beltrami operator (acting on the Hilbert space of the 
differential forms
defined on a Riemannian manifold): slight modifications of this 
example have been used to prove deep mathematical 
theorems~\cite{w1, bs}. 
By way of illustration, we now elaborate  briefly
on a quantum mechanical system whose
supersymmetric nature is not very familiar. 

\paragraph{The free particle in one dimension}
As pointed out quite recently~\cite{rau}, 
the free particle moving on a line represents the simplest
example of SUSYQM. 
In this case, the involution operator 
is realized~\cite{plyu}
by the 
{\em parity operator}$\,$\footnote{The author
of~\cite{plyu} refers to this choice as the `minimally 
bosonized SUSYQM' and discusses the one-dimensional
particle in a parity-odd superpotential.}:
\beq 
(K\varphi ) (x) = \varphi  (-x)
\qquad {\rm for} \ \; \varphi \in \hs = L^2 (\br )
\ .
\eeq 
Indeed, this operator is bounded, self-adjoint 
and satisfies $K^2 = \Id$. 
Since the momentum operator $p \equiv p_x$
changes sign under a parity transformation, 
$KpK^{\dagger} = -p$, we have $\{K,p \} =0$.
Henceforth, the operator $\Q = {1 \over \sqrt{2}} \, p$ represents
a supercharge for this quantum mechanical system: 
\beq
H = \Q^2 = {1 \over 2} \, p^2
\, , 
\qquad \{ K, \Q \} =0
\ .
\eeq

Let us verify that all of these expressions admit the standard matrix
representation. The decomposition (\ref{dsd}) 
into bosonic and fermionic vectors is now to be interpreted as a
decomposition into even and odd parity functions:
\begin{eqnarray*}
\varphi (x) \!\!\! & = & \!\!\!  {1\over 2} \, 
 [ \varphi (x) +\varphi (-x) ]
+ {1\over 2} \, [ \varphi (x) -\varphi (-x) ]
\\
\!\!\! & \equiv & \!\!\! \varphi_+ (x) + \varphi_- (x)
\ .
\end{eqnarray*}
The momentum operator modifies the parity,
\begin{eqnarray*}
p \varphi = (p \varphi )_+ + (p \varphi )_-
\!\!\! & = & \!\!\! 
{1 \over 2} \, ( \Id +K) \, p \varphi +
{1 \over 2} \, ( \Id -K) \, p \varphi
\\
\!\!\! & = & \!\!\! 
p \, {1 \over 2} \, ( \Id - K) \varphi +
p \, {1 \over 2} \, ( \Id + K) \varphi
\\
\!\!\! & = & \!\!\! 
p (\varphi_- ) + p (\varphi_+ )
\ , 
\end{eqnarray*}
i.e.
$(p \varphi )_{\pm} = p (\varphi_{\mp} )$.

Let us introduce the projection operators 
$\Pi_{\pm} = {1 \over 2} \, ( \Id \pm K)$ 
which satisfy $\Pi_+ \Pi_- = \Pi_- \Pi_+ =0$
and $\Pi_+ + \Pi_- = \Id$. Since $\{K,p \} =0$,
we obtain
\beq
\label{roi}
\left.
\begin{array}{l}
p = \Id p \Id = \Pi_- \, p \, \Pi_+ \, + \, \Pi_+ \,  p \, \Pi_-
\\
\qquad \qquad \!
= \ p \upharpoonright \hs_+ \; + \; p \upharpoonright \hs_-
\end{array}
\right\}
\quad {\rm with} \ \; 
\left\{
\begin{array}{l}
\Pi_- \, p \, \Pi_+ \, : \, \hs_+ \to \hs_- 
\\
\Pi_+ \,p \, \Pi_- \, : \, \hs_- \to \hs_+ 
\end{array}
\right.
\eeq
and $(\Pi_- \,p \, \Pi_+ )^{\dagger} = \Pi_+ \, p\, \Pi_-$.
For the sake of clarity, we presently put a hat on vectors 
and operators when referring to the matrix expressions:
\begin{eqnarray}
   \hs 
 \!\!\! & = & \!\!\!   
   \hs_+ \oplus \hs _- \ni 
  \left[
  \begin{array}{c}
\varphi_+ \\
\varphi_-
  \end{array}
  \right]
  \equiv \hat{\varphi}
  \, , \qquad 
\hat K =
 \left[
  \begin{array}{cc}
 \Id & 0 \\
0& - \Id
  \end{array}
  \right]
\nonumber
\\
\sqrt{2} \, \hat{\Q} \!\!\! & = & \!\!\!  \hat{p} \equiv
\left[
  \begin{array}{cc}
 0& 0 \\
p_{-+} & 0
  \end{array}
  \right]
+ 
\left[
  \begin{array}{cc}
 0& p_{+-} \\
0& 0
  \end{array}
  \right]
\label{restric}
\\
\hat H
\!\!\! & = & \!\!\! 
\hat{\Q} ^2
=
 {1 \over 2} \, 
 \left[
  \begin{array}{cc}
p_{+-} \, p_{-+}  & 0 \\
0& p_{-+} \, p_{+-}
  \end{array}
  \right]
 \, ,
\nonumber
\end{eqnarray}
where the two contributions of $\hat p$ correspond to those
of $p$ displayed in (\ref{roi}).

In this example, the spectrum of the superpartners 
$H_+ \equiv H \!  \! \upharpoonright \! \hs_+$ and 
$H_- \equiv  H \! \! \upharpoonright \! \hs_-$ is purely
continuous  and the 
generalized even and odd parity eigenfunctions  
associated to the spectral values $E= {1 \over 2} \, \rho ^2$
are given by 
\begin{eqnarray}
\varphi_{\rho}^{(+)} (x) \!\!\! & = & \!\!\! 
\cos \, (\rho x) \qquad {\rm for} \ \; \rho \geq 0
\nonumber \\
\varphi_{\rho}^{(-)} (x) \!\!\! & = & \!\!\! 
\sin \, ( \rho x) \qquad {\rm for} \ \; \rho>0
\, .
\end{eqnarray}
Thus,  $\rho =0$ is a non-degenerate 
spectral value
while the double degeneracy of all other 
spectral values
is a manifestation of 
supersymmetry~\cite{rau}.

By virtue of the result B stated in Subsection~\ref{summ}, 
a second supercharge exists and is given by 
$\Q_2 \equiv \pm \ri K \Q = {\pm \ri \over \sqrt 2} \, K p$.
However, this operator as well as the complex supercharge
${\bf q} = {1 \over \sqrt 2} \, (\Q + \ri \Q_2)$
are non-local since they explicitly involve
the parity operator.

\paragraph{The spin-${1 \over 2}$ particle
in a three-dimensional magnetic field}
A less basic example is given by Pauli's Hamiltonian
for a spin-${1 \over 2}$ particle
in a magnetic field $\vec B = \vec{{\rm rot}} \vec A$.
This operator acts on $\hs = L^2 (\br ^3) \otimes {\bf C}^2$
and has the form  
\begin{equation}
2 H=  (\vec p - \vec A \, ) ^2 \Id_2 - \vec B \cdot \vec{\sigma}
\, .
\end{equation} 
Here, $\vec{\sigma} = ( \sigma_1,\sigma_2,\sigma_3)$ 
and, for simplicity, we do not spell out the 
tensor product symbols in the present example. 
As was already noted in the early days of SUSYQM~\cite{crom},
we have 
\[
H= \Q^2 \, , \quad {\rm with} \ \; 
\sqrt{2} \, \Q = (\vec p - \vec A \, ) \cdot \vec{\sigma}
\]  
and $K \Q K^{-1} = -\Q$, where $K$ represents the parity 
operator\footnote{Note that 
$K\vec{\sigma}  K^{-1} = \vec{\sigma}$ since 
$\vec{S} \equiv {1 \over 2} \, \vec{\sigma}$ represents the 
angular momentum of spin.}.
The latter equation can also be rewritten as 
$\{ K, \Q \} =0$ and therefore we actually have a $n=1$ supersymmetric 
system. 
However, just as in the previous example, the second supercharge 
and the complex supercharge are given 
by non-local operators.

A simple matrix representation  for the state vector
$\phi = [ \varphi, \psi ]^t \in \hs$ 
(where `$t$' denotes transposition)
is defined 
by the $4$-component column vector
$\hat{\phi} = [ \varphi_+, \psi_+, \varphi_- , \psi_- ]^t$,
where $\varphi_+$ and
$\varphi_-$ denote, respectively,  
the even and odd parity 
parts of $\varphi \in L^2 (\br ^3)$.
The operators characterizing 
the supersymmetric system 
then read as 
\begin{eqnarray*}
\hat K
\!\!\! & = & \!\!\! 
 \left[
  \begin{array}{cc}
 \Id_2 & 0 \\
0& - \Id_2
  \end{array}
  \right]
  \, , \qquad 
  \sqrt{2} \, \hat{\Q} =
 \left[
  \begin{array}{cc}
 0& (\vec p - \vec A \, ) \cdot \vec{\sigma}   \\
(\vec p - \vec A \, ) \cdot \vec{\sigma} & 0
  \end{array}
  \right]
\\
\hat H
\!\!\! & = & \!\!\! 
\hat{\Q} ^2
=
 {1 \over 2} \, 
 \left[
  \begin{array}{cc}
(\vec p - \vec A \, ) ^2 \Id_2 - \vec B \cdot \vec{\sigma}  & 0 \\
0&  (\vec p - \vec A \, ) ^2 \Id_2 - \vec B \cdot \vec{\sigma}
  \end{array}
  \right]
\, .
\end{eqnarray*}
Here, we have suppressed the indices `$+$' and `$-$' denoting the 
restriction of operators to the subspaces $\hs_+$ and $\hs_-$,
see Equation (\ref{restric}).

\section{Construction of an involution 
from two supercharges}\label{relations}

In this section, we will deal with statement D 
made in Subsection~\ref{summ}, i.e., we will show that one 
can construct an involution operator $K$ from the two supercharges
$Q_1$ and
$Q_2$ defining a $N=2$ supersymmetric system. 
To do so, we first try to find a concrete expression for the 
involution which is present in a  $n=2$ 
supersymmetric system. 

As we have seen in Subsection \ref{gfn1}, 
the supercharges $\Q_1$ and $\Q_2$ defining a $n=2$  system 
are related by  
\beq
 \label{chq}
       \Q_2= - \ri \, K\Q_1 
       \eeq
(or $\Q_2= +\ri \, K\Q_1$). 
Equation (\ref{chq})
can be solved for $K$, 
\beq
  K  =  \ri \, \Q_2 \Q_1^{-1}
 \quad {\rm on} \ \; ({\rm ker} \, \Q_1 )^{\perp}
 \, ,
  \label{kqq}
  \eeq
where $({\rm ker} \, \Q_1 )^{\perp}$ denotes the orthogonal 
complement of the subspace ${\rm ker} \, \Q_1$. 

In view of expression (\ref{kqq}), we 
introduce an involution  into
the setting of Definition 1 by {\em defining}
\beq
  K  = \ri \, Q_2 Q_1^{-1}
 \quad {\rm on} \ \; ({\rm ker} \, Q_1 )^{\perp}
 \, . 
  \label{defp}
  \eeq
Clearly, the extension of the operator $K$ to all of $\hs$  
requires further discussion.
Before dealing with this issue, 
we note that relation $Q_1 ^2 = Q_2^2$ 
for two self-adjoint operators $Q_1$ and $Q_2$
implies that the kernels of these operators coincide: 
   \begin{equation}
  \label{qiq}
  Q_1 \varphi = 0 \ \Longleftrightarrow \   Q_2 \varphi = 0
    \ .
     \end{equation}     
In fact, $ Q_1 \vf =0$ is equivalent to 
\[
0 = \| Q_1 \vf \|^2 = \left\langle Q_1 \vf ,Q_1 \vf \right\rangle
=  \left\langle \vf ,Q^2_1 \vf \right\rangle 
=  \left\langle \vf ,Q^2_2 \vf \right\rangle 
 = \| Q_2 \vf \|^2
 \ ,
 \]
hence $ Q_2 \vf =0$.

Furthermore, for any  self-adjoint operator $Q$, we have 
the equivalence
  \begin{equation}
  \label{qq2}
  Q \varphi = 0 
  \ \Longleftrightarrow \   
  Q^2 \varphi = 0
    \ .
     \end{equation}   
Indeed, the left-hand-side 
obviously implies the right-hand-side
and the converse statement 
follows from 
\[
0 = \left\langle  \vf ,Q^2 \vf \right\rangle 
= \left\langle  Q \vf ,Q \vf \right\rangle 
= \| Q \vf \|^2 
\ .
 \]

   \begin{enumerate}\item[]
{\bf Theorem : }
     {\it 
Let $(H, \hs)$ be a supersymmetric system in the sense of 
Definition 1
with $N=2$, 
that is, assume there exist self-adjoint operators $Q_1$ and
$Q_2$ satisfying 
$Q_1^2 = Q_2^2 \equiv H$ and $\{ Q_1, Q_2 \} =0$.
Then, we have: 

{\rm (i)} 
The operator $K$ defined on $( {\rm ker} \, Q_1)^{\perp}$
by (\ref{defp}) admits 
  an extension to all of $\hs$ which represents a non-trivial 
  involution anticommuting with $Q_1$ and $Q_2$. 
  Hence, $(H, \hs)$ is a $n=2$  supersymmetric system in the sense of 
Definition 3.

{\rm (ii)} 
More specifically, 
if $Q_1$ has a kernel of finite dimension, then 
there exists a one-parameter family of 
extensions parametrized by the integer number 
${\rm ind_S} \, H \equiv d_+ - d_- $ where
$d_+ , d_- \in \{ 0,1,2, \dots \}$ are subject to the condition
\begin{equation}
\label{dpm}
d_+ + d_- = d \equiv {\rm dim \, ker} \; Q_1
   \ .
\end{equation}
Accordingly, there are 
$d+1$  
possible values for ${\rm ind_S} \, H$:
\[
{\rm ind_S} \, H \in \{ -d, -d+2, \dots, d-2, d \}
\, .
\]
}
   \end{enumerate}

   \noindent 
   {\bf Proof:} 
  The Hilbert space  $\hs$ can be decomposed into a direct sum  
   of the kernel of $Q_1$ (i.e., the eigenspace associated to 
   the eigenvalue zero) and its orthogonal complement: 
\beq
\label{dch}
   \hs = \hs_{0} \oplus \hs_{\perp}
    \qquad 
    {\rm with} \ \; 
    \hs_{0} = {\rm ker} \, Q_1
 = {\rm ker} \, H 
\, , \quad 
\hs_{\perp} = ( {\rm ker} \, Q_1)^{\perp}
 \ .
\eeq
 The operator $Q_1 \!  \upharpoonright \! \hs_{\perp}$ 
    is  invertible, hence   
    $K\! \upharpoonright \! \hs_{\perp} $ can be defined as in
    Eq.(\ref{defp}).  On the space $\hs_{\perp}$, the operator
  $K$ satisfies 
   \begin{equation}
      \label{basr}
      K^{\dagger} = K \, , \quad 
    K^2 = \Id  \;  , \quad 
\{ K, Q_1 \} = \{ K, Q_2 \} =0 
 \, .
      \end{equation}
    
    The definition (\ref{defp}) of $K$   on $\hs_{\perp}$
    is equivalent to the relation 
    \begin{equation}
\label{rev}
Q_2 = -\ri K Q_1 \qquad {\rm on} \ \; \hs_{\perp}
    \end{equation}     
    and the question is whether or not one can 
define an extension of the operator 
$K\! \upharpoonright \! \hs_{\perp} $ to all of $\hs$ 
such  that the relation $Q_2 = -\ri K Q_1$
and relations (\ref{basr}) also hold on 
$\hs_{0}$.  
The validity of $Q_2 = -\ri K Q_1$ 
on $\hs_{0}$  is equivalent to 
  \[
 Q_2 \vf = - \ri \, K Q_1\vf 
 \qquad \mbox{for all} \ \; \vf  \in \hs_{0}
 \ . 
 \]
 This equation holds trivially since 
 both sides vanish by virtue of (\ref{qiq}),  
  whatever the expression of $K$. 
  Thus, the only constraints
  for the definition of $K$ on 
$\hs_{0}$ consist of the conditions 
  $K^{\dagger} = K$ and $K^2= \Id$. 
Operators with these properties exist and any one 
of them will be suitable 
for our theorem. In particular, 
if the eigenvalue zero of $Q_1$ is of finite multiplicity, 
then  $ K\!  \upharpoonright \! \hs_{0}$ is (up to unitary equivalence) 
a  diagonal matrix with eigenvalues $\pm 1$, i.e.,
     \[
   K\!  \upharpoonright \! \hs_{0} = {\rm diag} \, 
   \left( \, 1,\dots , 1\, ; \, -1, \dots ,-1
   \, \right) \ , 
   \]
     with $d_+ \geq 0$ entries $1$ and  $d_-\geq 0$ entries $-1$,  
    subject to the condition (\ref{dpm}).
Since there is no constraint for the integer
$d_+ - d_- $,
any value satisfying 
$|d_+ - d_- | \leq d_+ + d_- \equiv d$
can be chosen: each one gives rise to another extension
of $K$ to all of $\hs$.

 \section{Concluding remarks}

Our discussion shows that a precise answer to 
the question raised in the title of our paper
can only be given if one specifies the definition of 
SUSYQM that one has in mind. Obviously, 
$n=1$ and $n=2$ SUSYQM are equivalent.  
Furthermore, 
it seems that the examples presented in the literature
for $N=1$ SUSYQM (like Pauli's Hamiltonian) 
actually represent examples of 
$n=1$ since an involution exists,  
though the second supercharge of the ensuing 
$n=2$ system is non-local in this case.

There is no reason for 
a true $N=1$ supersymmetric system (i.e., $H=Q_1^2$
with no involution $K$ that satisfies $\{ K, Q_1 \} =0$)
to be equivalent to a $N=2$ system. For such a system, 
the property of being `even' or `odd' is not defined 
and therefore one cannot infer either any of 
the typical properties associated with supersymmetric systems. 
Accordingly, the statement made in the pioneering work~\cite{w1}
that 
``the simplest supersymmetric quantum mechanical system
has $N=2$'' should indeed be interpreted as 
saying that $N=1$ SUSYQM is not truly supersymmetric.
 
From our discussion, we can conclude that 
the conceptually simplest approach to SUSYQM
is the approach that starts with one supercharge $\Q$
that anticommutes with 
an 
involution operator $K$.
Simpler or physically more transparent 
expressions can eventually be obtained by
using $\Q$ and $K$ to construct a second supercharge or 
by introducing a complex supercharge in terms of the latter
two charges.

The line of arguments presented in our work
can be generalized to  a large extent
to the case of SUSYQM with more than two supercharges, 
as well as to 
parasupersymmetric and fractional supersymmetric quantum mechanics. 
This discussion is beyond the scope of the present letter and 
will be reported upon elsewhere~\cite{prepar, cgk}.

\bigskip 
\medskip

\noindent{\bf Acknowledgments :} 
F.G. is grateful to D.~Maison 
for a valuable discussion which was at the origin of 
Section~\ref{relations}.
He also wishes to thank the organizers of the 
conference {\em ``Progress in supersymmetric quantum
mechanics''} for the stimulating atmosphere and for their 
support.  

\bigskip



\begin{thebibliography}{22}
\newcommand{\artref}[5]{{\sc #1}: {\it #2}, #3 {\bf #4} #5}
\newcommand{\bookref}[3]{{\sc #1}: ``{\it #2}$\,$", #3}
\newcommand{\prepref}[3]{{\sc #1}: {\it #2}, #3}

  
\bibitem{nic}
\artref
{H.~Nicolai}{Supersymmetry and spin 
systems}{J.Phys.}{A9}{(1976) 1497.}
 
\bibitem{w1}
\artref
{E.~Witten}{Dynamical breaking of 
supersymmetry}{Nucl.Phys.}{B188}{(1981) 513;}

\artref
{E.~Witten}{Constraints on supersymmetry 
breaking}{Nucl.Phys.}{B202}{(1982) 253;}

\artref
{E.~Witten}{Supersymmetry and Morse
theory}{J.Differ.Geometry}{17}{(1982) 661.}



\bibitem{mqs}
\bookref
{F.~Cooper, A.~Khare and U.~Sukhatme}{Supersymmetry in Quantum
Mechanics}{(World Scientific, 2001);}

\bookref
{B.K.~Bagchi}{Supersymmetry in Quantum and Classical 
Mechanics}{(Chapman and Hall, 2001);}


\bookref
{G.~Junker}{Supersymmetric Methods in Quantum 
and Statistical Physics}{(Springer Verlag, 1996).}
 
 
\bibitem{bs}
\bookref
{H.L.~Cycon, R.G.~Froese, W.~Kirsch and B.~Simon}{Schr\"odinger
Ope\-rators - with Applications to Quantum Mechanics and Global
Geometry}{(Springer Verlag, 1987).}
 

\bibitem{vasil}
\artref
{F.-H.~Vasilescu}{Anticommuting self-adjoint 
operators}{Rev.Roum.Math.Pures et Appl.}{28}{(1983) 77;}

\artref
{S.~Pedersen}{Anticommuting self-adjoint 
operators}{J.Funct.Anal.}{89}{(1990) 428;}

\bookref
{A.~Arai}{Analysis on anticommuting self-adjoint operators}{in 
``Advanced Studies in Pure Mathematics 23, Spectral and Scattering Theory
and Applications'', ed. K.Yajima  (1994).}



\bibitem{wess}
\bookref
{J.~Wess and J.~Bagger}{Supersymmetry and 
Supergravity}{(Princeton University Press, 1983).}


\bibitem{berezin}
\bookref
{F.A. Berezin}{Introduction to Superanalysis}{Mathematical
Physics and Applied Mathematics Vol.9
(D.Reidel Publ. Co., Dordrecht 1987).}



\bibitem{tha}
\bookref
{B.~Thaller}{The Dirac Equation}{(Springer Verlag, 1992).}


\bibitem{rau}
\bookref
{A.~Rau}{Supersymmetry 
in quantum mechanics: An extended view}{preprint 
LG9208, and talk presented at the Conference 
{\em ``Progress in supersymmetric quantum
mechanics''} (Valladolid, 15-19 July 2003).}


\bibitem{plyu}
\artref
{M.~Plyushchay}{Supersymmetries in pure 
parabosonic systems}{Int.J.Mod.Phys.}{A15}{ (2000) 3679.}
 
\bibitem{crom}
\artref
{M.~de Crombrugghe and V.~Rittenberg}{Supersymmetric quantum mechanics}{
Annals Phys.}{151}{(1983) 99.}

\bibitem{prepar}
\bookref
{F.Gieres and M.~Lefran\c cois}{The 
general structure of
supersymmetric quantum mechanics 
(and of some of its generalizations)}{in preparation.}


\bibitem{cgk}
{\sc M.~Combescure and F.Gieres,}{$\,$ work in progress.} 




\end{thebibliography}
\end{document}